\documentclass[a4paper,fleqn,usenatbib]{mnras}

\usepackage{newtxtext,newtxmath}

\usepackage[T1]{fontenc}
\usepackage{ae,aecompl}

\usepackage{graphicx}	
\usepackage{amsmath}	
\usepackage{amssymb}	
\usepackage{enumerate}
\usepackage{multicol}
\usepackage{multirow}
\usepackage{enumerate}
\usepackage{ulem}
\usepackage{xcolor}
\usepackage{soul}

\title[Future MCAO astrometric observations]{The impact of geometric distortions in multiconjugate adaptive optics astrometric observations with future extremely large telescopes}

\author[M. Patti et al.]{
M. Patti$^{1}$\thanks{E-mail: mauro.patti@inaf.it}, 
C. Arcidiacono$^{2}$, M. Lombini$^{1}$, E. Diolaiti$^{1}$, F. Cortecchia$^{1}$
\\
$^{1}$INAF - Osservatorio di Astrofisica e Scienza dello Spazio di Bologna, Italy\\
$^{2}$INAF - Osservatorio Astronomico di Padova , Italy\\
}

\date{Accepted 2019 May 06. Received 2019 April 23; in original form 2019 February 21}

\pubyear{2019}

\begin{document}
\label{firstpage}
\pagerange{\pageref{firstpage}--\pageref{lastpage}}
\maketitle

\begin{abstract}
Astrometry is one of the main science case which drives the requirements of the next multiconjugate adaptive optics (MCAO) systems for future extremely large telescopes. The small diffraction limited point-spread function (PSF) and the high Signal-to-Noise Ratio (SNR) of these instruments, promise astrometric precision at the level of micro-arcseconds. However, optical distortions have to be as low as possible to achieve the high demanding astrometry requirements. In addition to static distortions, the opto-mechanical instabilities cause astrometric errors that can be major contributors to the astrometry error budget. The present article describes the analysis, at design level, of the effects of opto-mechanical instabilities when coupled with optical surface irregularities due to the manufacturing process. We analyse the notable example of the Multi-conjugate Adaptive Optics RelaY (MAORY) for the extremely large telescope (ELT). Ray-tracing simulations combined with a Monte Carlo approach are used to estimate the geometrical structure and magnitude of field distortion resulting from the optical design. We consider the effects of distortion on the MCAO correction showing that it is possible achieve the micro-arcseconds astrometric precision once corresponding accuracy is obtained by both optical design and manufacturing. We predict that for single-epoch observations, an astrometric error below 50 $\mu as$ can be achieved for exposure times up to 2 min, provided  about 100 stars are available to remove fifth-order distortions. Such performance could be reproducible for multi-epoch observations despite the time-variable distortion induced by instrument instabilities. 
\end{abstract}




\section{Introduction}
The future large telescopes aim to enhance the astrometric performance for ground-based near-infrared (NIR) instruments. The Extremely Large Telescope (ELT)~\citep{tamai2018eso}, the Thirty Meter Telescope (TMT)~\citep{sanders2013thirty} and the Giant Magellan Telescope (GMT)~\citep{johns2008giant} will achieve resolution almost five or four times better than current 8-10 meter class telescopes~\citep{mountain1994gemini, iye2004current, matthews1996first}. The requirements of these giant instruments set the astrometric accuracy to the level of tens of micro-arcseconds for relative astrometry on single and multi-epoch observations~\citep{pott2018micado, schock2016flowdown}.\\ Multiconjugate adaptive optics (MCAO)~\citep{beckers89a,beckers88} is appropriate for astrometry because it provides atmospheric turbulence correction over a relative large field of view (FoV), i.e.$\sim$1 arcmin$^2$. The science will benefit of uniform and high-quality point-spread functions (PSFs) over the FoV reducing the centroiding errors.\\
The PSF centroiding precision depends on the Signal to Noise Ratio (SNR) and can be analytically derived using diffraction theory in the photon-noise regime~\citep{lindegren1978photoelectric}:
\begin{equation}
\sigma_{x,y}=\frac{1}{\pi}\:\frac{\lambda}{D\cdot\sqrt{N}}\approx\frac{1}{\pi}\:\frac{FWHM}{SNR}
\label{eq:1}
\end{equation}
$N$ is the number of collected photons by a telescope of diameter $D$. The Full Width Half Maximum (FWHM) of the diffraction limited PSF is equal to $1.03\lambda / D$ which is different from the radius of the first null in an Airy disk~\citep{airy1835diffraction}.\\ With a large collecting area, it is theoretically possible to achieve the goal of 10 micro-arcsecond precision, e.g. during a typical ELT image exposure time (e.g. 1min) in $H$ band with SNR of the order of $250-300$. However, numerous errors degrade the astrometric observations. These are related to the instrument instabilities, optical distortion, astronomical errors and atmospheric effects~\citep{schock2014thirty}~\citep{trippe2010high}.\\
This article focuses on the instrumental errors due to opto-mechanical instabilities and optics manufacturing, modelling the behaviour of MAORY during MCAO operations, as a meaningful example.
At the ELT, each NIR image is mainly affected by plate scale distortions due to gravity-induced flexures~\citep{rodeghiero2018impact}. The impact of plate scale variations can be controlled with the MCAO correction but distortions, higher then plate scale, could be an issue and have to be studied carefully. The analysis, here presented, is based on ray-tracing simulations combined with a Monte Carlo approach. The main purpose of the article is to describe the methodology applied to the study of geometric distortion in order to predict the astrometric performance of future instruments. The analysis is purely optical and does not consider the atmospheric turbulence. 
A similar ray-tracing study has recently been performed for the Gemini MCAO system  ~\citep{10.1093/mnras/stz596}. The results are consistent not only with that coming from use of a calibration mask~\citep{riechert2018gems}, but also with distortion corrections performed directly on the sky. This means that the distortion measurements of previous works ~\citep{massari2016astrometry, dalessandro2016gems, bernard2018optimal} were not affected by systematic errors and other sources of distortions (like atmospheric/AO terms) were negligible with respect to the optical distortions.
In this context, an accurate optical design model could be useful to predict, with good approximation, the astrometric performances of future instruments. The analysis, based on a ray-tracing model, is applied to MAORY to illustrate its case.
After a brief description of MAORY design, Section~\ref{sec:3} describes the two metrics used to evaluate the astrometric precision and quantifies the nominal instrument optical distortion on the basis of the MAORY astrometric requirement. Section~\ref{sec:4} describes the impact of distortion on the MCAO correction. Section~\ref{sec:5} addresses the effects of instrument instabilities and Section~\ref{sec:6}, the optics manufacturing errors. In Section~\ref{sec:7}, instrument instabilities are coupled with optics manufacturing errors and a multi-epoch astrometric analysis conclude the article.

\section{MAORY}
\label{sec:2}
\begin{figure}
\centering
	\includegraphics[width=0.85\columnwidth]{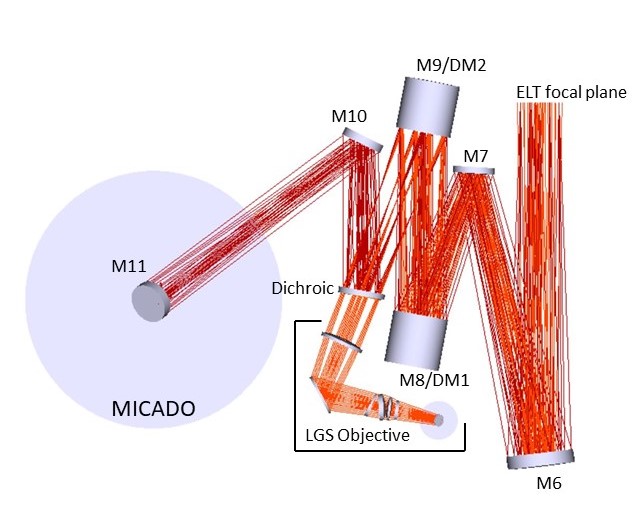}
    \caption{MAORY optical design layout}
    \label{fig:1}
\end{figure}

MAORY~\citep{diolaiti2016maory} is the MCAO module of the ELT. Its wave-front (WF) sensing architecture is based on six Laser Guide Stars (LGSs) and three Natural Guide Stars (NGSs). The high-order WF sensing is performed by using LGSs while the low-order WF sensing is performed by using NGSs to measure the modes which cannot be accurately sensed by the LGSs. The optical design of MAORY, here considered, is based on six mirrors (including one or two deformable mirrors (DMs)) and a dichroic beam-splitter. The dichroic reflects the NIR light to the NGS wave-front sensors (WFSs) and the science instruments and transmits the LGS light to the LGS WFSs by means of an objective~\citep{lombini2018optical}.  
In MCAO mode, MAORY is required to provide an unvignetted corrected field of diameter $D\approx75$ arcsec for the multi-adaptive optics imaging camera for deep observations (MICADO)~\citep{davies2018micado} while the MCAO correction is performed over the field of diameter $D\le180$ arcsec.
The three NGSs are approximately detected within the area enclosed between the MICADO FoV and the technical FoV radius of $R_{tec}\le90$ arcsec. Within this patrol field, three probes (movable pick-off mirrors) will gather the light from three suitable NGSs to perform the low-order WF sensing. The NGS WFSs are in common path with the MICADO entrance focal plane through the so-called Green Doughnut~\citep{bonaglia2018status} which is the interface between MICADO cryostat and MAORY exit focal plane. The MAORY optical design layout~\citep{10.1093/mnras/stz810} is shown in Figure~\ref{fig:1} and it is composed by:
\begin{itemize}
\item The Main Path Optics (MPO) is a 1:1 (in terms of F-ratio) optical relay with mirrors labelled from M6 (the first mirror in MAORY, after M5 in the telescope) to M11. All mirrors have optical power except M11 which is a flat folding mirror for the MICADO gravity invariant port.
\item The LGS objective is a F/5 mostly refractive optical design and it projects the images of artificial stars for each of the six LGS WFS. The laser beacon generates the stars at a typical altitude of 90 $Km$, modulated by both the variation of sodium layer altitude and the elevation perspective effect in a range between 80 $km$ and 240 $km$.
\end{itemize}
%
The following analysis is focused on the MPO which delivers the light to the science instruments. Mirrors with optical power are pure conic surfaces. M10 is the only mirror with even aspheric terms. The optical surface parameters are listed in Table~\ref{tab:baseline}. Unlike the other mirrors with optical power, the DMs are not an off-axis section from a larger parent surface. They have a similar curvature radius and small aspheric deviation: less than 3$\mu m$ peak to valley. The different surface asphericity was assumed to be achieved by the actuator themselves.
\begin{table}
\centering
\caption{Optical surface parameters of the MPO design. Concave surfaces are labelled as \textit{cv}, convex surfaces as \textit{cx}.}
	\label{tab:baseline} 
	\renewcommand{\tabcolsep}{2pt}
\begin{tabular}{lccrcr}
\hline
\multicolumn{1}{l}{\textbf{Surfac}e} & \multicolumn{1}{l}{\begin{tabular}[c]{@{}l@{}}\textbf{Decentre}\\  (mm)\end{tabular}} & \multicolumn{1}{l}{\begin{tabular}[c]{@{}l@{}}\textbf{Tilt}\\ (deg)\end{tabular}} & \multicolumn{1}{l}{\begin{tabular}[c]{@{}l@{}}\textbf{Curv.} \\ \textbf{radius}\\  (mm)\end{tabular}} & \multicolumn{1}{l}{\begin{tabular}[c]{@{}l@{}}\textbf{Conic} \\ \textbf{const.}\end{tabular}} & \multicolumn{1}{l}{\begin{tabular}[c]{@{}l@{}}\textbf{Aspheric}\\ \textbf{terms}\end{tabular}} \\ \hline
M6 & 399 & 6 & 11923 \textit{(cv)} & -1.8 & - \\
M7 & 770 & 17 & 6044 \textit{(cx)} & -1.0 & - \\
M8/DM & 0 & 10 & 13880 \textit{(cv)} & -5.3 & - \\
M9/DM & 0 & 10 & 13901 \textit{(cv)} & 2.2 & - \\
Dichroic & - & 11 & Infinity & - & - \\
M10 & 469 & 27 & 53906 \textit{(cx)} & 0.0 & \begin{tabular}[c]{@{}r@{}}4th:7.2e-13\\ 6th:-2.8e-19\\ 8th:1.5e-25\end{tabular} \\
M11 & - & 45 & Infinity & - & -
\end{tabular}
\end{table}
\section{Optical distortion analysis}
\label{sec:3}
\begin{figure}
	\includegraphics[width=\columnwidth]{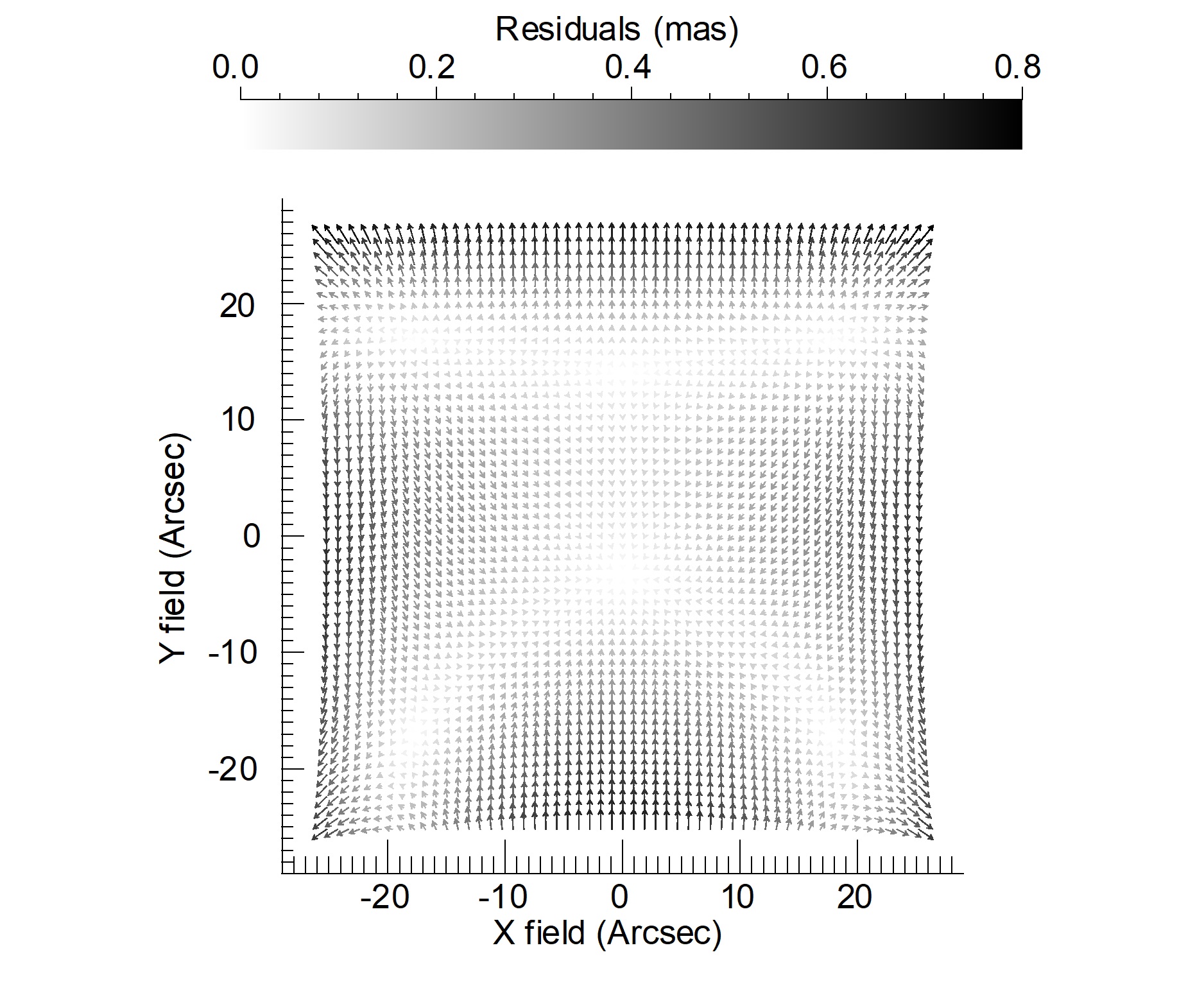}
    \caption{Distortion map of the MAORY optical design. These are contributions of second or higher orders. The 53''$\times$53'' FoV is considered where the vectors lenghts have been scaled to highlight the distortion structures}
    \label{fig:2}
\end{figure}

\begin{figure}
\centering
	\includegraphics[width=0.9\columnwidth]{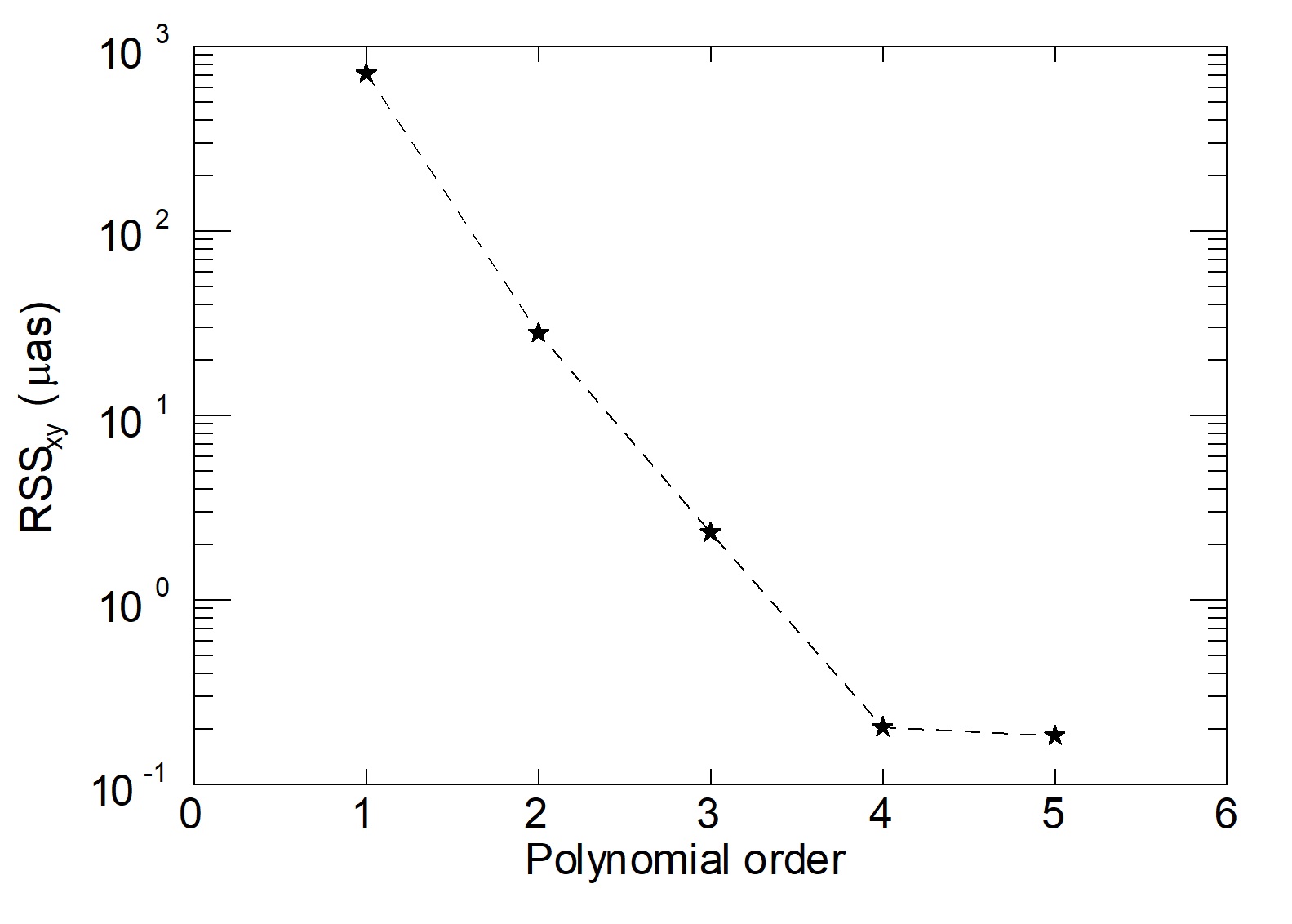}
    \caption{Astrometric residual errors as defined in equation~\ref{eq:3} after fitting a polynomial of order $1\leq n\leq5$. It refers to the nominal optical system without degradation from manufacturing and alignment errors; nominal telescope design included.}
    \label{fig:4}
\end{figure}

\begin{figure}
\centering
	\includegraphics[width=0.8\columnwidth]{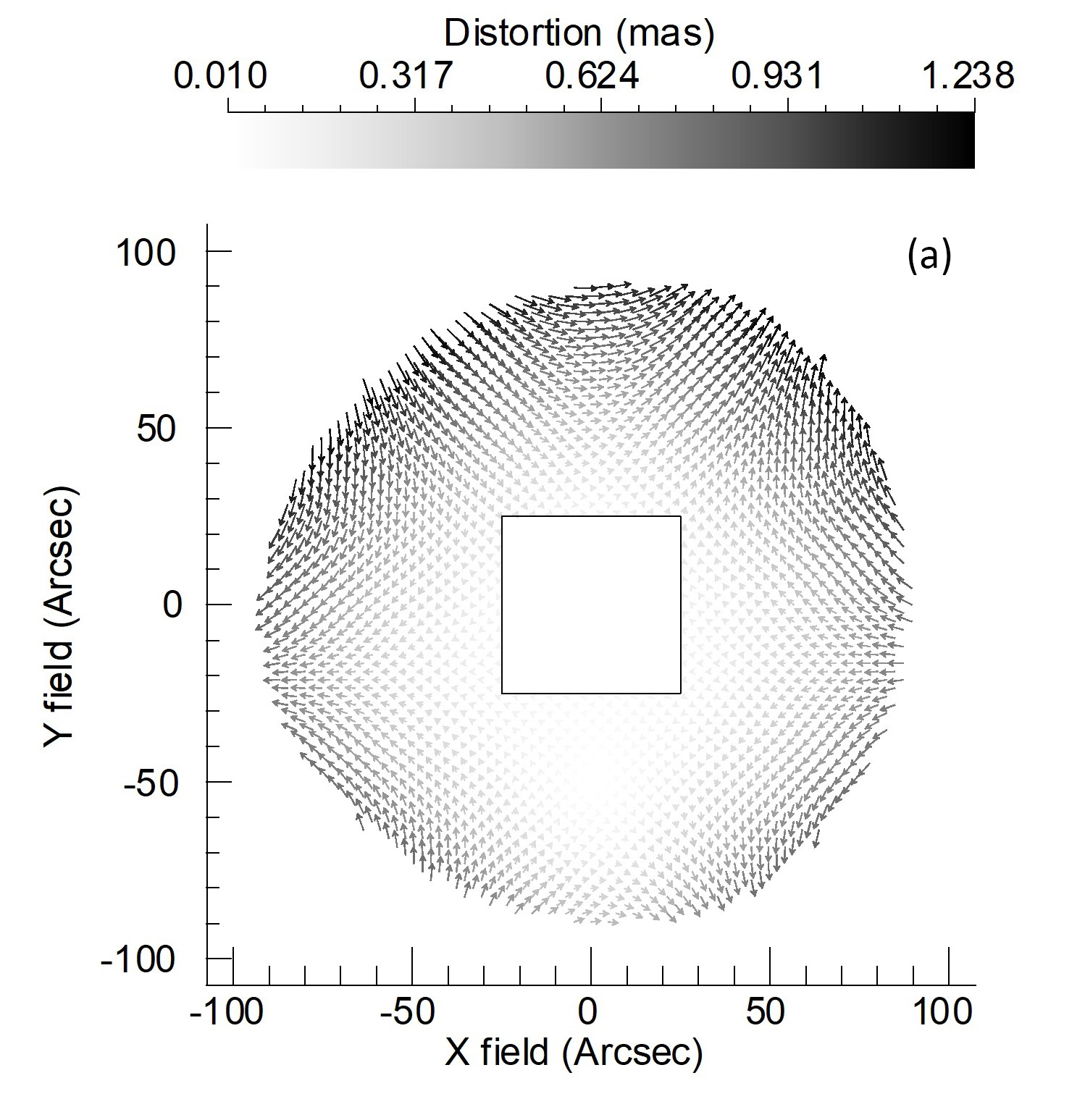}\vspace{2em}
	\includegraphics[width=0.8\columnwidth]{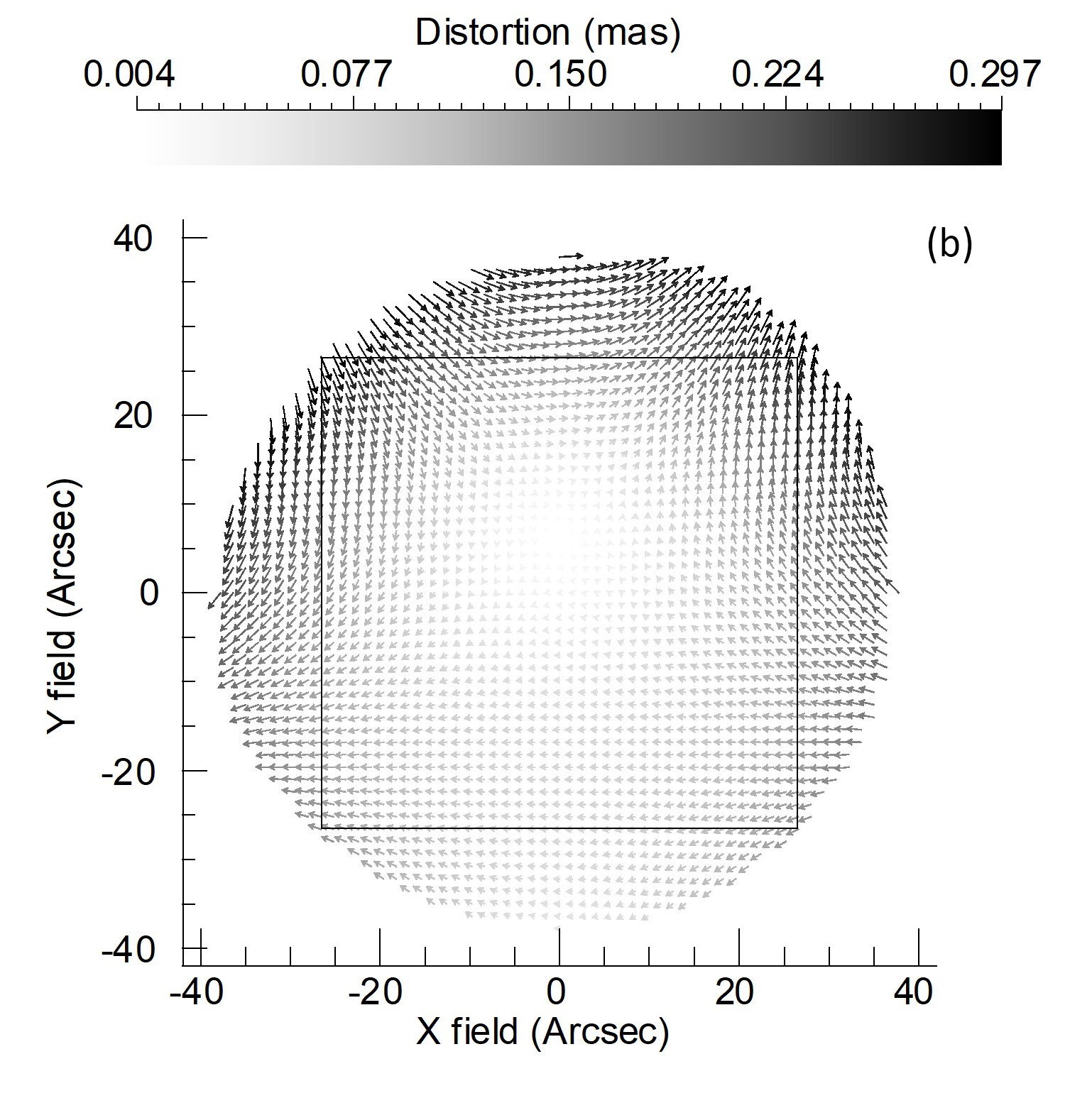}
    \caption{PSF drift at the MAORY exit port within the maximum integration time for narrow band astrometric observations (without degradation from manufacturing and alignment errors; nominal telescope design included). (a): PSF drift in the NGS patrol FoV. (b): PSF drift in the circle containing the MICADO FoV (black square).}
    \label{fig:3}
\end{figure}

The instrument design phase is extremely important to minimize system distortions that could worsen the astrometric accuracy. Ray-tracing softwares, such as Zemax\textsuperscript{\textregistered}, allow to simulate the instrument behaviour under different conditions which affect the final optical quality. A complete optical analysis of the instrument performance~\citep{patti2018maory} could be obtained by means of user-defined scripts that allow to control and perform simulations from the ray-tracing software environment.\\ We use a set of reference sources in the so-called object space. The PSFs centroids of these sources are detected in the image space after a sequential ray-tracing from sources through optics to focal plane. The comparison of the coordinates list of the reference sources with the corresponding PSF centroid coordinates in the image space, returns the quantitative measure of the distortions. In order to match the reference catalog coordinates with the detector coordinates, a linear transformation has to be applied to take into account rigid shifts, rotations and different scale. For this reason, only contributions of second or higher order to the astrometric error are usually called distortions~\citep{massari2016astrometry}. Given a FoV, the term `relative astrometry' refers to measurement of different science objects separations relative to each other or relative to other field objects. These measurements have to be converted in sky coordinates to be scientifically useful and that's the aim of `absolute astrometry'. The MAORY distortion map is shown in Figure~\ref{fig:2} over the 53''$\times$53'' scientific FoV. 
\\
Astrometry is one of the primary science case for MICADO and MAORY shall permit observations that reach astrometric precision of 50 $\mu as$ with the goal of 10 $\mu as$~\citep{pott2018micado}.
\\As previously anticipated, given a set of science images, the detector positions have to be converted into global astrometric coordinates. This operation aims to solve the instrument distortion by matching a reference catalog coordinates with the detector coordinates.\\
The typical calibration procedure adopted in astrometric observations is that any two images, or the coordinates of the relevant sources in the two images, are transformed onto each other by a polynomial transformation of $n^{th}$ order. It is common to use a least squares fit where the independent variables are the reference star coordinates $(x_{r},y_{r})$ while the dependent variables are distorted star coordinates $(x_{d},y_{d})$. The following analyses use the polynomials defined as:
\begin{equation}
\label{eq:2}
\begin{split}
x_{f} = \sum^n_{i=0}\sum^n_{j=0}\:K^{i,j}_x\:x^i_{r}\:y^j_{r}\\
y_{f} = \sum^n_{i=0}\sum^n_{j=0}\:K^{i,j}_y\:x^i_{r}\:y^j_{r}
\end{split}
\end{equation}
The polynomial order $n$ indicates the maximum of the sum of the exponents $i$ and $j$ in the equations. The $K^{i,j}_x$ and $K^{i,j}_y$ are $(n+1)$ square matrixes used to model the distortions. The residual distortion map is given by $(x_{r}-x_{f})$, $(y_{r}-y_{f})$ and the astrometric residual error is defined as:
\begin{equation}
\label{eq:3}
RSS_{xy} = \sqrt{\sigma^2_{x}+\sigma^2_{y}}
\end{equation}
It is the Root Sum of Squares (RSS) of $\sigma_{x}$, $\sigma_{y}$ specified as the $x,y$ standard deviations of the residual distortion map and it is shown in Figure~\ref{fig:4}. This is the lowest achievable astrometric error since it refers to the nominal optical system and there are no measurement errors. In fact, the $RSS_{xy}$ is evaluated in a static scenario where the centroid coordinates on the detector are known with 'infinite' precision. In a realistic scenario, the $RSS_{xy}$ precision is affected by the dynamics within a single exposure image (integration time). Our analysis does not consider the atmosphere and the nominal optical design has diffraction-limited performance in the NIR region.
\\In Figure~\ref{fig:4}, as well as in the following Sections where $RSS_{xy}$ results are shown, about 100 point-like sources have been used to fit the polynomials from first to fifth order distortions. The astrometric observations with ELTs relies on on-sky calibrations and there might be regions in the sky where such calibration is not possible due to the lack of standard astrometric stars. ~\cite{rodeghiero2018impact} estimated the number of expected Gaia~\citep{prusti2016gaia} reference stars over 1 arcmin$^2$ FoV to be used for distortion calibration. Furthermore, they also reported the minimum number of stars with suitable SNR for different orders of polynomial fit, stating that only in crowded regions of a globular cluster, the number of Gaia stars will be largely sufficient to correct third order distortions. We underline that the stars have to be uniformly distributed over the FoV to achieve the lowest astrometric error related to the considered degree of polynomial fit. The correction of fifth or higher order distortions could be addressed by means of self-calibration~\citep{ anderson2000toward} or it could be possible when the completeness of the Gaia survey will improve.\\ MAORY will be placed on the ELT Nasmyth platform and its optics are not co-rotating with the sky during the telescope tracking. If a field de-rotation is not applied at the focal plane, the PSF of a point-like source follows a trajectory which may be described as an arc of a circle. The MPO, which relay the telescope focal plane to the exit port for the science instrument, is affected by asymmetric distortions (see Figure~\ref{fig:2}). This means the PSF circular trajectory is affected by small perturbations due to distortions during the telescope tracking. Field de-rotation only compensates for the circular part of the trajectory so that any deviations from a pure rotation could degrade the PSF. Setting the SNR, the $RSS_{xy}$ precision will depend on the PSFs FWHM of the point-like sources (equation~\ref{eq:1}) used to perform the polynomial fit. The astrometric precision will decrease if the PSFs move from their positions, due to the geometric distortion, at the beginning of the exposure time.
The PSF drift requirement of the MPO derives from the astrometric requirement of MAORY/MICADO. For narrow band astrometric observations, MICADO aims to achieve a SNR of the order of 250-300 using reasonably bright stars of magnitude $H = 19-23$ and 120 seconds as maximum single exposure time. $H$ band astrometry is preferred since it is the working wavelength of the NGS WFSs which measures tip-tilt aberrations, so it looks the optimal choice to minimise differential atmospheric chromatic effects between the WFSs and MICADO. During the maximum single exposure time, MAORY should not introduce any drift that makes the PSF FWHM larger than 1/10 of its diffraction limit size. Such limit corresponds to 0.88 $mas$ confining the FWHM to 9.7 $mas$ at wavelength $\lambda=1.6\: \mu m$. \\Under these assumptions, strictly applying equation~\ref{eq:1} with $SNR=300$, it is possible to achieve the 10 $\mu as$ centroiding precision. Considering the $K$ band, the requirement on the PSF FWHM enlargement is relaxed to 1.2 $mas$ ($\lambda=2.2\: \mu m$).\\ The PSF drifts are an error source of the polynomial fit and in the end, the astrometric precision is the root-sum-squared of the $RSS_{xy}$ and the centroiding precision due to the PSF drift. This is a conservative approach to evaluate the instrument astrometric performances which allows to separate the two contributions described above.
\\Here we quantify the PSF drift, setting the integration time to the maximum one foreseen for narrow band astrometric observations of $T=120\:s$. The telescope elevation is fixed at 80$^\circ$ (i.e. 10$^\circ$ from zenith) and, given the ELT site coordinates, it is possible to derive:
\begin{itemize}
\item $A\approx 13.7 \times (1/\cos(80^\circ)) \approx 79\:arcsec/s$. It is the maximum de-rotator angular velocity;
\item $T \times A \approx 2.6^\circ$. It is the maximum field rotation within a single exposure image.
\end{itemize}
Taking the maximum field rotation as a reference, it is possible to constrain the maximum integration time as follows ($h=$ telescope elevation angle):
\begin{itemize}
\item $T = 120 s, \:h = 80^\circ$;
\item $T = 60 s, \: h = 85^\circ$;
\item $T = 30 s, \:h = 87^\circ$.
\end{itemize}
In general, for a given elevation angle, the shorter the exposure time, the lower the PSF enlargement due to optical distortion. \\
The positions of a set of $N$ test stars placed over a regular grid have been used to evaluate the PSF drift. We define $X_{1}$, $Y_{1}$, the initial coordinates of star centroids and $X_{2}$, $Y_{2}$ the coordinates of the star centroids after the maximum astrometric exposure time (i.e. 2.6$^\circ$ FoV rotation and counter-rotation angle). The PSF centroids move, due to distortions, by $\Delta X=(X_{2}-X_{1})$ and $\Delta Y=(Y_{2}-Y_{1})$. The vector sum of $\Delta X$ and $\Delta Y$ is the PSF drift experienced by the $n^{th}$ test star and shown in Figure~\ref{fig:3}. It refers to the MAORY nominal design (nominal telescope design included) without degradation from manufacturing and alignment errors. These drifts are added in quadrature to the diffraction limited FWHM, implying that 4 $mas$ is the limit, in terms of centroid shift, which cause the 0.88 $mas$ of FWHM enlargement at $\lambda=1.6\: \mu m$. Considering $\lambda=2.2\: \mu m$, this limit is relaxed to 5.5 $mas$.\\ The PSF drifts, in the case of the nominal design, are largely within the limit, as can be seen in Figure~\ref{fig:3} (b). The maximum value is about 0.3 $mas$ at the edge of the scientific FoV. When added in quadrature to the diffraction limited FWHM in $H$ band (8.8 $mas$), it introduces a negligible FWHM enlargement. Strictly applying equation~\ref{eq:1} with $SNR=300$, the MAORY nominal design achieve a centroiding precision less than 10 $\mu as$. When added in quadrature to the third-order $RSS_{xy}$ which is about 2.5 $\mu as$ (Figure~\ref{fig:4}), the instrument is able to achieve the goal of 10 $\mu as$ of astrometric precision.
%
\section{Impact of distortions on MCAO}
\label{sec:4}
If during the MCAO operation, the WFSs detect aberrations which are not related to the atmosphere but originated by the distortion pattern, the applied correction is not optimal and a further degradation of the PSF may be introduced.
The NGS low order MCAO loop, through the asterism of three NGSs, will be affected mainly by two effects related to the geometric distortion:
\begin{itemize}
\item	$0^{th}$ order errors: Tip-Tilt errors, related to the average NGSs position;
\item	$1^{st}$ order errors: plate scale errors, generated by the change of relative distances of the NGSs.
\end{itemize}
In a possible scenario, the MCAO operation concept plans to drive the WFSs probes to the apparent star location which includes atmospheric refraction and optical distortions (a potential initial position fine-tuning could be necessary). After closing the loop, the stars are kept in place on the NGS probes by the field de-rotator and the telescope tracking. If the FoV is affected by optical distortion, as discussed in the previous Section, the WFSs will detect a differential tip-tilt (shift) of the NGS constellation respect to the nominal on-sky position. This could translates into plate scale distortions that will affect and potentially destroy the astrometric solution. There are three plate scale modes: plate scale stretched in $X$ and $Y$, respectively, and shear. Thus, if three or more reference objects are available in the MICADO frame, it is possible to use them to calibrate the plate scale error. Conversely, if the NGSs are used as references, a differential motion among the probes will introduce a corresponding plate scale error in the MICADO field. In a general sense, these errors could be defined as positioning errors and have to be properly calibrated (e.g. by look up tables).\\ 
To evaluate the impact of NGSs distortions on the MICADO FoV, we opted for a Monte Carlo approach, by the calculation of random asterisms of three NGSs. The stars are placed at the vertices of random triangles and shall satisfy the following conditions:
\begin{itemize}
\item	The three NGSs must not be co-linear;
\item	The NGS coordinates must be included in the area delimited by the scientific FoV and the patrol FoV $\leq$180 arcsec;
\item	The minor length of the asterism side is, at least, half the scientific FoV.
\end{itemize}
Let $(X_{1},\: Y_{1})$ be the initial coordinates of the stars centroids, $(X_{2},\: Y_{2})$ the coordinates of the stars due to distortions after the field rotation and counter-rotation within the single astrometric image. The errors have been modelled as vectors:
\begin{equation}
\begin{split}
X_{2}&=A_{x}+B_{x}\cdot X_{1}\\
Y_{2}&=A_{y}+B_{y}\cdot Y_{1} 		
\end{split}
\label{eq:fit}
\end{equation}                                                                  
The coefficients $(A_{x},\:A_{y})$ are the global tip-tilt components of the geometric distortion, while $(B_{x}, B_{y})$ are the $1^{st}$ order errors due to differential tip-tilt between the probes (plate scale). The algorithm used to evaluate the $A_{x,y}$ and $B_{x,y}$ coefficients works as follows:
\begin{enumerate}
\item We suppose to place the probes at the NGS focal plane positions, at the centroid coordinates $(X_{1},\: Y_{1})$.
\item The asterism rotates by 2.6$^\circ$ (see Section~\ref{sec:3}) around the optical axis being fixed on sky. We assume that MICADO field de-rotator continuously counter-rotates the field.
\item The new centroid coordinates $(X_{2},\: Y_{2})$ don't overlap with the initial ones because of the optical distortions.
\item	The same procedure (points (ii) - (iii)) is applied to a grid of test stars within the MICADO FoV. This is the nominal PSF drift as shown in Figure~\ref{fig:3}.
\item	Given the centroids $(X_{1},\: Y_{1})$ and $(X_{2},\: Y_{2})$, we solve the equations~\ref{eq:fit} to find the coefficients $A_{x,y},\:B_{x,y}$.
\item	Let $(X^\prime_{1},\: Y^\prime_{1})$ be the coordinates of a regular grid of stars on sky within the MICADO FoV. Use the solutions find in point (v) to calculate the distorted coordinates  $(X^\prime_{2},\: Y^\prime_{2})$.
\item	The MCAO-induced distortion onto the MICADO FoV are given by  $\Delta X^\prime=-(X^\prime_{2}-X^\prime_{1})$  and $\Delta Y^\prime=-(Y^\prime_{2}-Y^\prime_{1})$.
\item	The final PSF drift within the MICADO FoV is given by $\Delta X +\Delta X^\prime$ and $\Delta Y+\Delta Y^\prime$. 
\item	Repeat the points from (i) to (iii) and from (v) to (viii) considering different asterisms, each one characterized by different coefficients $A_{x,y},\:B_{x,y}$ and thus, different distortions.
\end{enumerate} 
\begin{figure}
\centering
	\includegraphics[width=0.9\columnwidth]{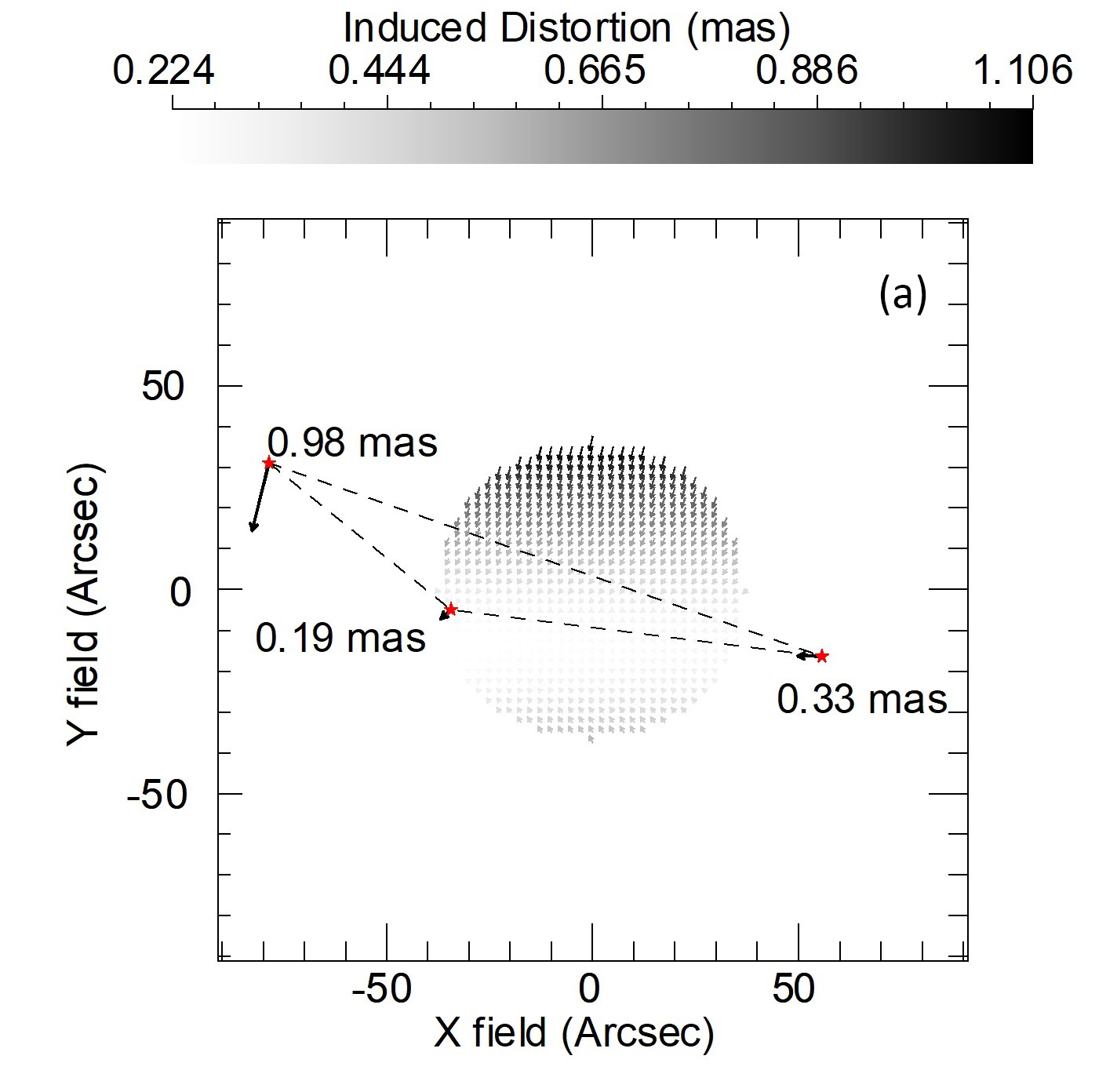}\vspace{2em}
		\includegraphics[width=0.9\columnwidth]{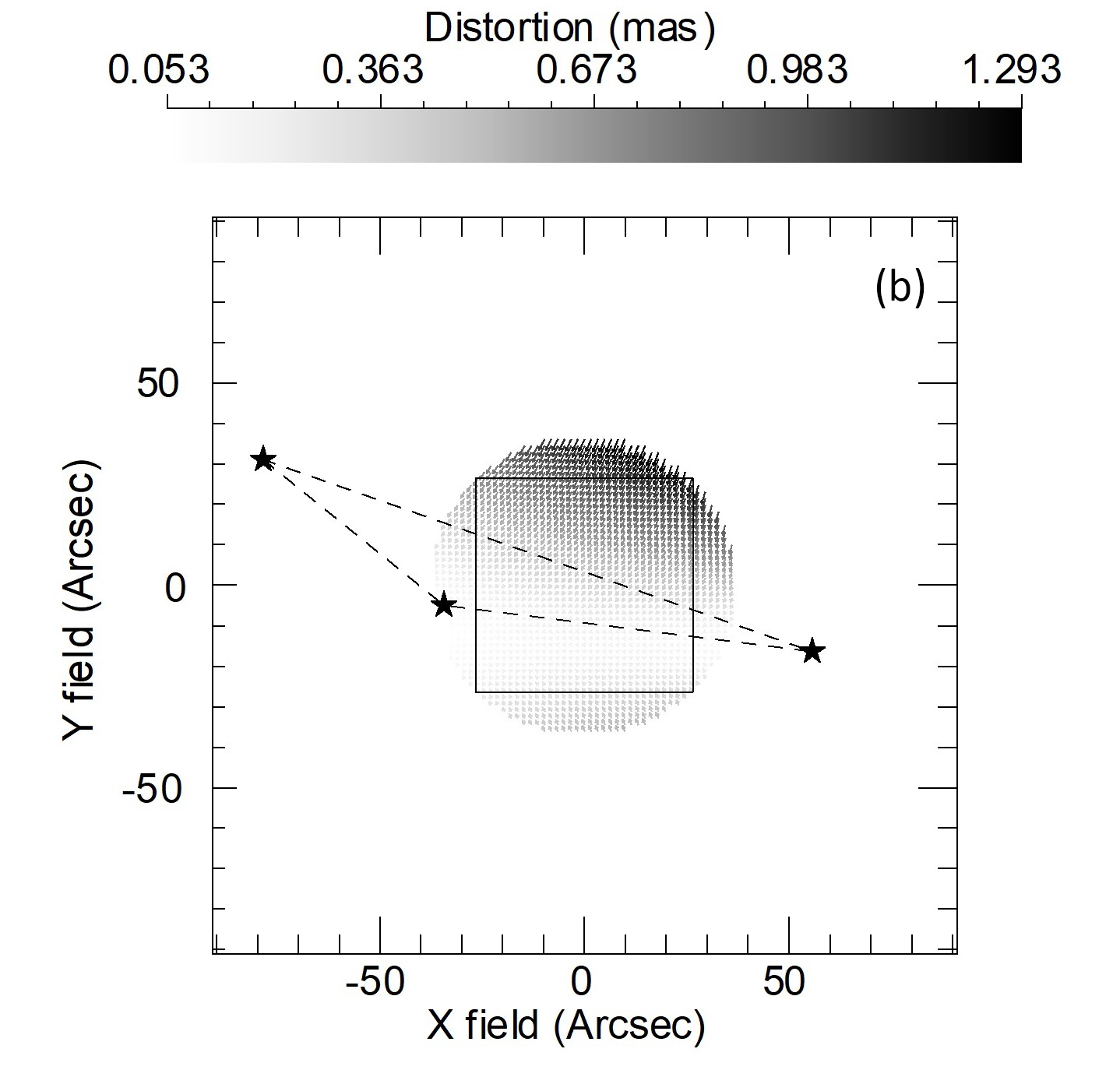}
    \caption{(a): Distortion map induced to the MICADO FoV (75'' diameter) by Tip-Tilt and plate scale errors for a NGS asterism considered as example. The PSF drift of each NGS composing the asterism is labelled in the Figure. (b): Vector map of PSF drift in the MICADO FoV for a single exposure image after adding the contribution of the NGS asterism (i.e. algebraic sum of NGS induced distortions and MICADO FoV PSF drift).}
    \label{fig:5}
\end{figure}
%
\begin{figure}
\centering
	\includegraphics[width=0.9\columnwidth]{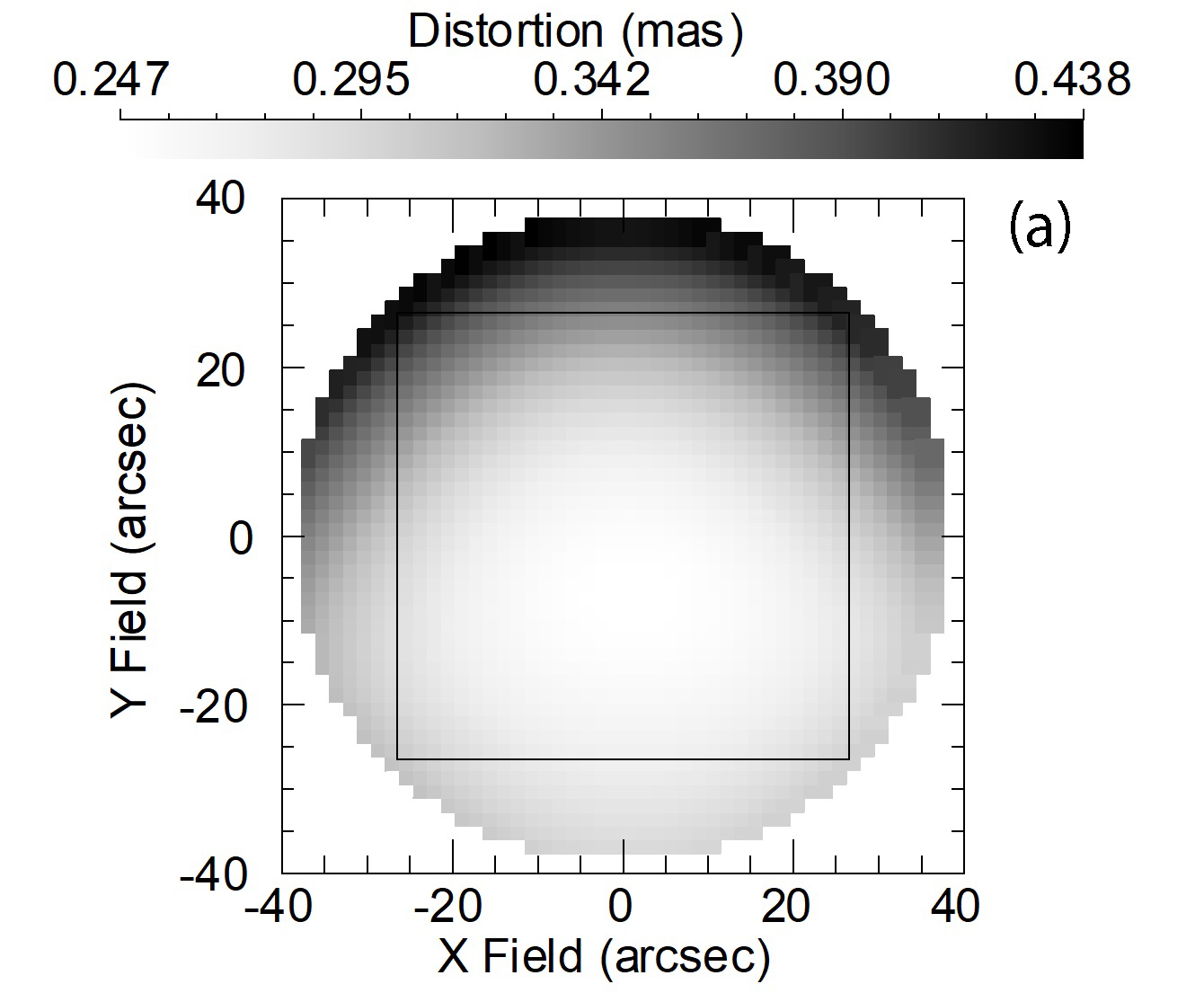}
	\includegraphics[width=0.75\columnwidth]{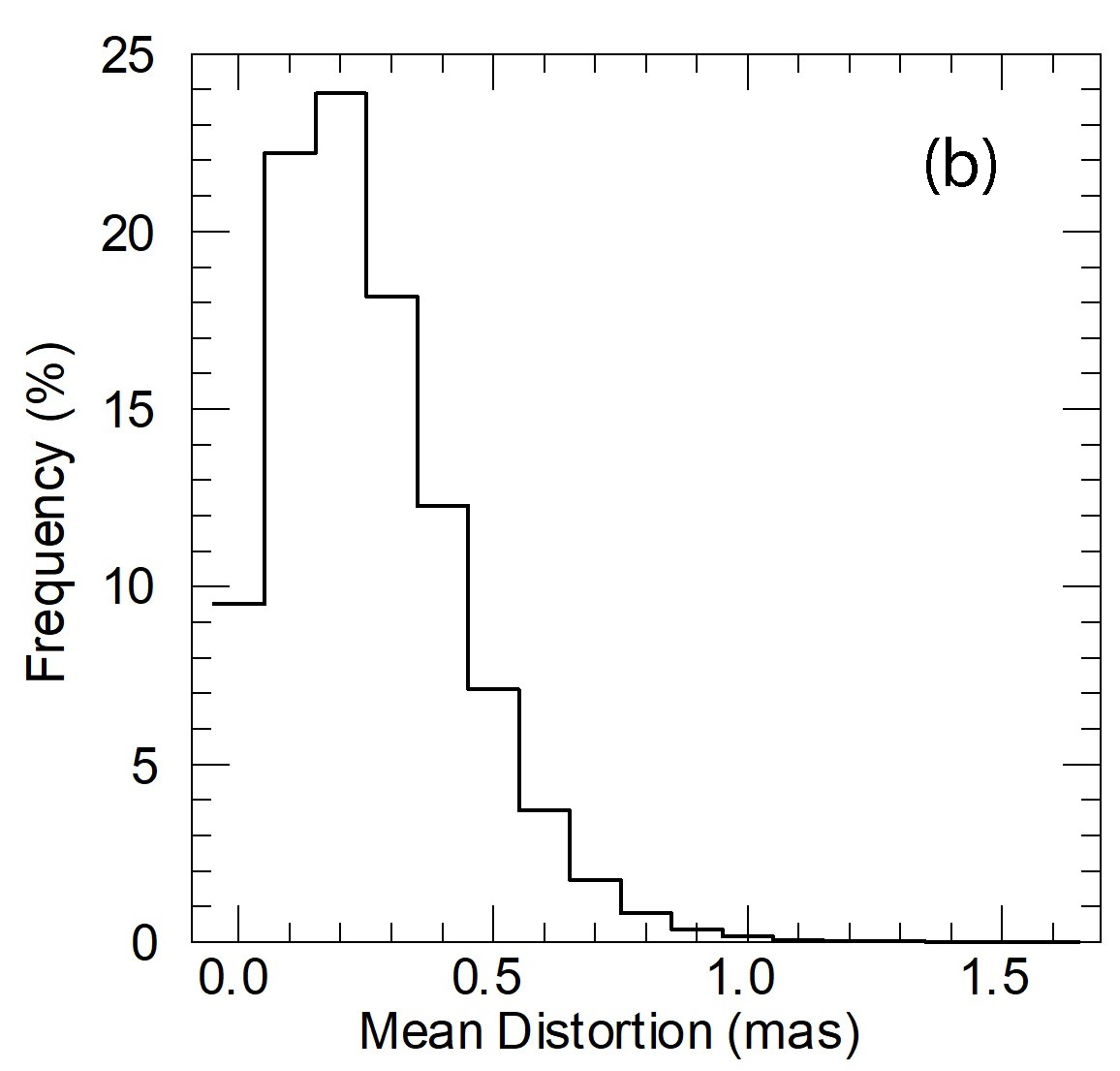}
    \caption{(a): Map of the average amplitudes of the PSF drifts over the trials for one astrometric image in the MICADO FoV. We obtained this result by adding the optical distortions on the science FoV and the errors propagated through the MCAO and induced by the optical distortions in the technical FoV. (b): Distribution of all the PSF drifts values for one astrometric image in the MICADO FoV as obtained for all the asterism realizations.}
    \label{fig:6}
\end{figure}
Figure~\ref{fig:5} shows one result of the algorithm described above. It is the worst NGS asterim, in terms of PSF drift, that we extracted from the Monte Carlo trials. The mean contribution over 10000 random realizations and the distribution of the resulting PSF drifts are shown in Figure~\ref{fig:6}. Typical drifts are smaller than 0.5 $mas$ with a mode of 0.2 $mas$. These values are well below the requirement of 4 $mas$ to have a centroiding precision of 10 $\mu as$ and thus, to achieve the astrometric precision in terms of $RSS_{xy}$ post-fit error (Figure~\ref{fig:4}). In the analysis, only the nominal distortions from the optical design are considered. Actually, other effect may super-impose to the considered optical distortion: i.e. the thermal stability of the probe mechanics, or flexures due to the rotation of the gravity vector. However, we are considering timescales $\leq$2 minutes, so these effects should be negligible.
\section{Effect of instrument instabilities}
\label{sec:5}
Instabilities of the MCAO system (telescope + instrument) have the potential to affect astrometric precision. Considering the light beam coming from a source, instabilities cause the beam footprint to wander on the optics generating distortion map variations.\\This Section concentrates only on the astrometric errors generated by the instrument instabilities, trying to model the behaviour of MAORY in typical operation scenarios.\\The analysis of ELT instabilities has been made by~\cite{rodeghiero2018impact} who found distortions variations up to $\approx 5 mas/arcmin$ to be dominated by plate scale and third order terms. Moreover, the authors assert these distortion variations can be suppressed significantly by two DMs operated in a MCAO system.\\In the MAORY analysis, a similar approach has been used. Expected instabilities of the optics have been injected into the nominal optical design in order to evaluate their impact on the distortion pattern variations and the residual astrometric errors. The DMs are considered to be rigid mirrors in order to decouple the instrument effects from the MCAO loop. The expected positioning errors comes from the tolerance analysis~\citep{patti2018maory} where a range of tolerable instrument instabilities has been defined considering an error budget for instrument performance~\citep{diolaiti2011elt}. The MPO opto-mechanical tolerances are listed in Table~\ref{tab:1}, these include deformations of the bench within one observing night and mounting stability tolerances. They have been used to generate 500 random optical system realizations which meet the specified range with a parabolic distribution. This is a conservative approach for the generation of possible expected performance since all applicable tolerances are simultaneously, and more likely, considered at the extreme ends of the tolerance range. After fitting a polynomial of order $1\leq n\leq5$ as defined in equation~\ref{eq:2}, the astrometric residual error due to tolerable instrument instabilities has been computed and it is shown in Table~\ref{tab:insta} as median and standard deviation of the Monte Carlo simulation. The $RSS_{xy}$ standard deviation is more indicative about the impact of instabilities on the performance since the median $RSS_{xy}$ value is very close to the corresponding nominal $RSS_{xy}$ shown in Figure~\ref{fig:4}. It is clear from Table~\ref{tab:insta} that instrument instabilities are confined to the third-order distortion correction.
\begin{table}
\centering
\caption{ Required stability of the MPO and the MAORY optical bench. We consider the MAORY module as rigid body, and specify the tolerances with respect to the ELT optical axis. The values represent the required stability after system alignment. }
\label{tab:1}
\begin{tabular}{ll}
\hline
\multicolumn{2}{c}{\textbf{Internal tolerances stability}} \\ \hline
\begin{tabular}[c]{@{}l@{}}X-Y Tilts \\   (M7 - M9 - M10 - Dichroic)\end{tabular} & $\pm 30\:\mu rad$ \\ \hline
\begin{tabular}[c]{@{}l@{}}X-Y Tilts \\ (M6)\end{tabular} & $\pm 87.5\:\mu rad$ \\ \hline
\begin{tabular}[c]{@{}l@{}}X-Y Tilts \\ (M8 - M11)\end{tabular} & $\pm 17.5\:\mu rad$ \\ \hline
\begin{tabular}[c]{@{}l@{}}Z Tilt\\ (All optics rotation around optical axis)\end{tabular} & $\pm 30\:\mu rad$ \\ \hline
\begin{tabular}[c]{@{}l@{}}X-Y Decenters \\ (All optics)\end{tabular} & $\pm 0.1\:mm$ \\ \hline
\begin{tabular}[c]{@{}l@{}}Z (axial) Decenter \\ (All optics)\end{tabular} & $\pm 1\:mm$ \\ \hline
\multicolumn{2}{c}{\textbf{MAORY module stability (rigid body)}} \\ \hline
X-Y-Z Tilts & $\pm0.25 mrad$ \\
X-Y Decenters & $\pm2.5 mm$ \\
Z (axial) Decenter & $\pm5 mm$ \\ \hline
\end{tabular}
\end{table}
%
%
\begin{table}
\centering
\caption{$RSS_{xy}$ is the astrometric residual errors as defined in equation~\ref{eq:3} after fitting a polynomial of order $1\leq n\leq5$. The table lists the median and standard deviation of $RSS_{xy}$ distribution as result of the Monte Carlo simulation due to tolerances listed in Table~\ref{tab:1}.}
\label{tab:insta}
\begin{tabular}{ccc}
\hline
\multicolumn{3}{c}{\textbf{$RSS_{xy}$ Monte Carlo statistic}} \\ \hline
Polynomial order & Median ($\mu as$) & $\sigma$ ($\mu as$) \\ \hline
1 & 712 & 25 \\
2 & 27.96 & 0.68 \\
3 & 2.54 & 0.07 \\
4 & 0.174 & 0.003 \\
5 & 0.169 & 0.003 \\ \hline
\end{tabular}
\end{table}
Operating MAORY in the MCAO mode and thus, correcting the Tip-Tilt and plate scale variations (first-order distortion correction), the expected astrometric precision is of the order of tens of micro-arcsec (see Table~\ref{tab:insta}), achieving the sub micro-arcsec precision after a third order polynomial fit.\\ 
The PSF drift, as described in Section~\ref{sec:4}, due to instrument instabilities, is negligible since most of the distortion variations are Tip-Tilt and plate scale errors and these low order aberrations are well corrected by the MCAO loop. As in the case of the nominal design, when added in quadrature to the third-order $RSS_{xy}$ which is about 2.5 $\mu as$ (Table~\ref{tab:insta}), the instrument is able to achieve the goal of 10 $\mu as$ of astrometric precision. However, a realistic scenario of operative instrument must include not only time-variable distortions due to opto-mechanical instabilities but also a model of manufactured optical surfaces. This is the topic of next Section where  astrometric performance are also considered in a multi-epoch scenario. 
\section{Effect of optics manufacturing}
\label{sec:6}
\begin{figure}
	\includegraphics[width=\columnwidth]{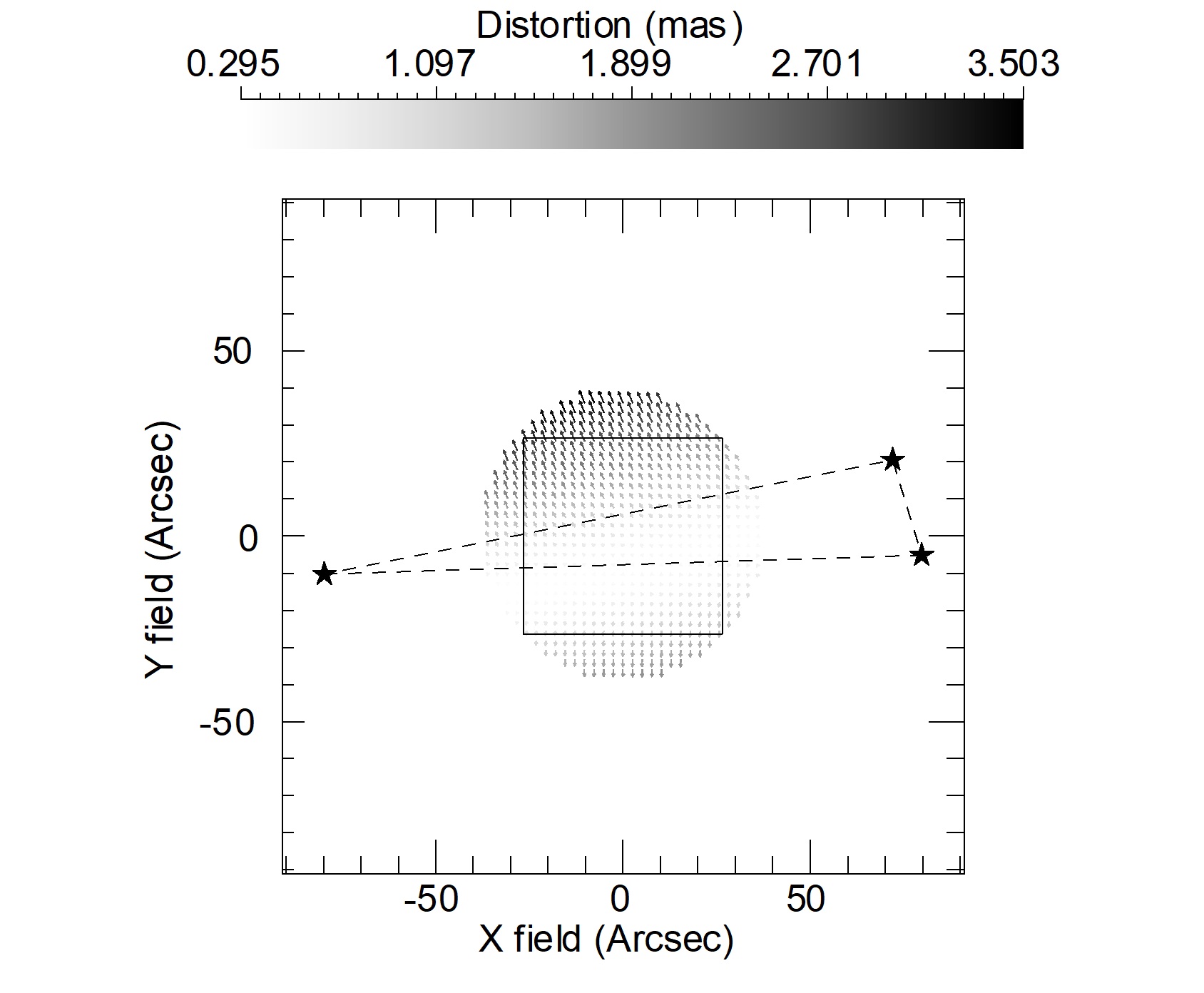}
    \caption{Same of Figure~\ref{fig:5}(b) but considering the worst Monte Carlo trial in terms of distortions after manufacturing.}
    \label{fig:8}
\end{figure}

\begin{figure}
\centering
  \includegraphics[width=0.9\columnwidth]{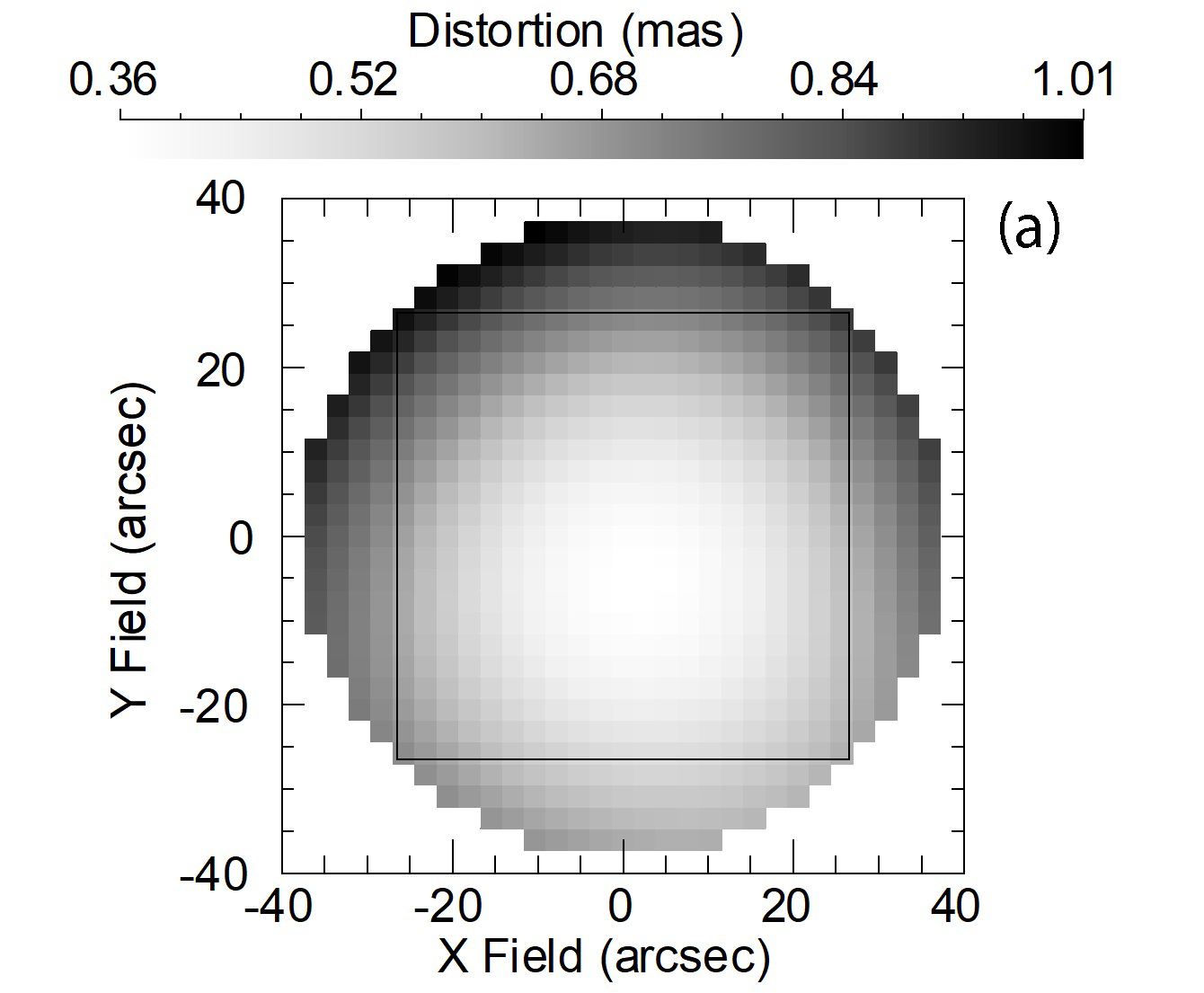}
	\includegraphics[width=0.75\columnwidth]{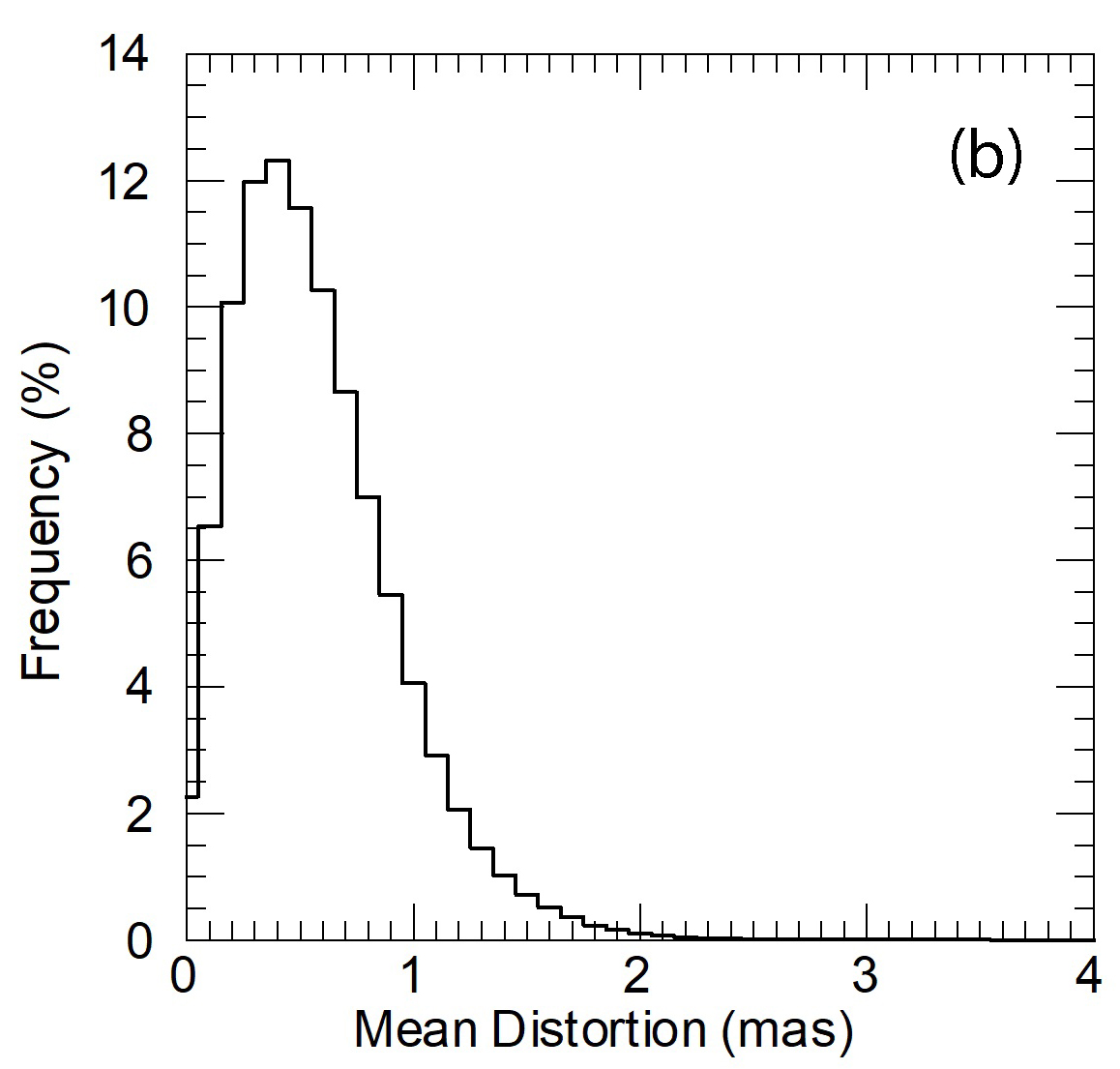}
    \caption{Same of Figure~\ref{fig:6} but considering the worst Monte Carlo trial in terms of distortions after manufacturing.}
    \label{fig:9}
\end{figure}

Optics manufacturing is not a perfect process. The optical surface will always deviate from its nominal shape in a range which is set by design through tolerance analysis. The MAORY optics manufacturing tolerances are summarized in Table~\ref{tab:2}~\citep{patti2018maory}. Surface irregularities can be described as waviness with a given period and amplitude~\citep{erdei2004tolerancing}. In general, the surface diameter sets the upper limit on the period when half cycle of waviness is over the whole surface affecting the nominal WF. The beam footprint diameter sets the lower limit on the period when more than one cycle of waviness is over the beam footprint and surface irregularities causes scattered light instead of additional WF error (WFE). This is related to the surface roughness which is not considered in the analysis. To take care of waviness period and amplitude, surface irregularities are modelled by means of standard Zernike coefficients~\citep{noll1976zernike} divided into ranges whose maximum tolerance value is the exact RMS error of the surface.
Tolerance on Zernike coefficient 11 (Z11, the "spherical" term) is considered separately because it is equivalent to a variation on the surface conic constant~\citep{patti2018maory}. For flat surfaces (dichroic and M11), the curvature tolerance is expressed in terms of distance on the normal from the surface to the center of the curvature (`Sagitta'). The number of Zernike coefficients that Zemax\textsuperscript{\textregistered} can simulate is not infinite. However, the values of the last Zernike range could be interpreted as Z$>$120 for the purpose of a manufacturing process.\\
Considering different field points, the beams footprints overlap differently on each optical surface. As the field rotates during the observations, the beams footprints cross the static irregularities introducing a dynamical centroid variation. The lower is the overlap of different footprints, the bigger is the difference in terms of irregularities crossed by a given beam. This means that surfaces with the largest conjugate ranges introduce the largest error since the farther the surface is from the pupil plane, the smaller is the optical footprint on that surface. In these terms, the flat folding mirror (M11) is the worst offender for MAORY.
\begin{table}
\centering
\caption{MPO manufacturing tolerances. Irregularities of optical surfaces are divided into consecutive ranges of Zernike modes. The amplitude values are the exact RMS error of the surface. $Z=11$ (spherical aberration) for flat surfaces is included in the budget of irregularities.}
\label{tab:2}
\renewcommand{\tabcolsep}{4.5pt}
\begin{tabular}{cccccc}
\hline
\multicolumn{6}{|c|}{\textbf{Manufacturing tolerances}} \\ \hline
Curvature radius & \multicolumn{5}{c}{$\pm$ 0.1 \%} \\
\begin{tabular}[c]{@{}c@{}}Sag of flat surfaces\\  (residual curvature)\end{tabular} & \multicolumn{5}{c}{$\pm$ 316 nm} \\ \hline
\multicolumn{6}{|c|}{\textbf{Spherical aberration (RMS error)}} \\ \hline
\begin{tabular}[c]{@{}c@{}}All mirrors with\\  optical power\end{tabular} & \multicolumn{5}{c}{5 nm} \\ \hline
\multicolumn{6}{|c|}{\textbf{Surface irregularity (RMS error)}} \\ \hline
\multicolumn{1}{l}{} & \multicolumn{5}{c}{Zernike range} \\ \hline
\multicolumn{1}{|c|}{Surface} & \multicolumn{1}{c|}{5-10} & \multicolumn{1}{c|}{12-28} & \multicolumn{1}{c|}{29-45} & \multicolumn{1}{c|}{46-120} & \multicolumn{1}{c|}{121-230} \\ \hline
M6 & 24 nm & 5 nm & 5 nm & 4 nm & 4 nm \\
M7 & 18 nm & 3 nm & 5 nm & 4 nm & 4 nm \\
M8 & 15 nm & 5 nm & 5 nm & 4 nm & 4 nm \\
M9 & 14 nm & 3 nm & 5 nm & 4 nm & 4 nm \\
M10 & 20 nm & 3 nm & 5 nm & 4 nm & 4 nm \\
M11 & 23 nm & 5 nm & 5 nm & 4 nm & 4 nm \\
Dichroic & 11 nm & 5 nm & 5 nm & 4 nm & 4 nm \\ \hline
\end{tabular}
\end{table}
\\
500 Monte Carlo trials have been considered to evaluate the impact of optics manufacturing listed Table~\ref{tab:2}. The $RSS_{xy}$ median and standard deviation of the Monte Carlo simulation are listed in Table~\ref{tab:manu}.\\
The effect on astrometry has been evaluated also in terms of PSF drift during single-epoch observations as defined in Section~\ref{sec:4}. The worst Monte Carlo trial, in terms of geometric distortion, has been taken as reference optical model. Random NGS asterims have been considered as described in Section~\ref{sec:4} and the results are shown in Figure~\ref{fig:8} and~\ref{fig:9}. The worst asterism case (Figure~\ref{fig:8}) introduces, at the edge of the science field, a PSF drift of 3.5 $mas$ while the median of PSF drifts distribution is about 1 $mas$. Typical drifts are smaller than 1 $mas$ with a mode of 0.5 $mas$ (see Figure~\ref{fig:9}). When 1 $mas$ is added in quadrature to the diffraction limited FWHM in $H$ band (8.8 $mas$), the FWHM enlargement still allows to achieve a centroiding precision less than 10 $\mu as$ with $SNR=300$ (equation~\ref{eq:1}). In this case, at least a fourth-order $RSS_{xy}$ (see Table~\ref{tab:manu}) is necessary to achieve the requirement of 50 $\mu as$ of astrometric precision as quadrature-sum of $\sim$10 $\mu as$ (centroid precision) and $\sim$30 $\mu as$ (Table~\ref{tab:manu}, 4$^{th}$ order). The goal of 10 $\mu as$ of astrometric precision is achieved in $\sim$10\% of the Monte Carlo trials with the fifth-order $RSS_{xy}$. 
\begin{table}
\centering
\caption{Same of Table~\ref{tab:insta} but the Monte Carlo simulation considers the tolerances listed in Table~\ref{tab:2}.}
\label{tab:manu}
\begin{tabular}{ccc}
\hline
\multicolumn{3}{c}{\textbf{$RSS_{xy}$ Monte Carlo statistic}} \\ \hline
Polynomial order & Median ($\mu as$) & $\sigma$ ($\mu as$) \\ \hline
1 & 831 & 109 \\
2 & 74.6 & 34.9 \\
3 & 50.4 & 15.7 \\
4 & 28.9 & 12.4 \\
5 & 13.5 & 3.9 \\ \hline
\end{tabular}
\end{table}
%
\section{Multi-epoch analysis}
\label{sec:7}
\begin{figure}
\centering
  \includegraphics[width=0.9\columnwidth]{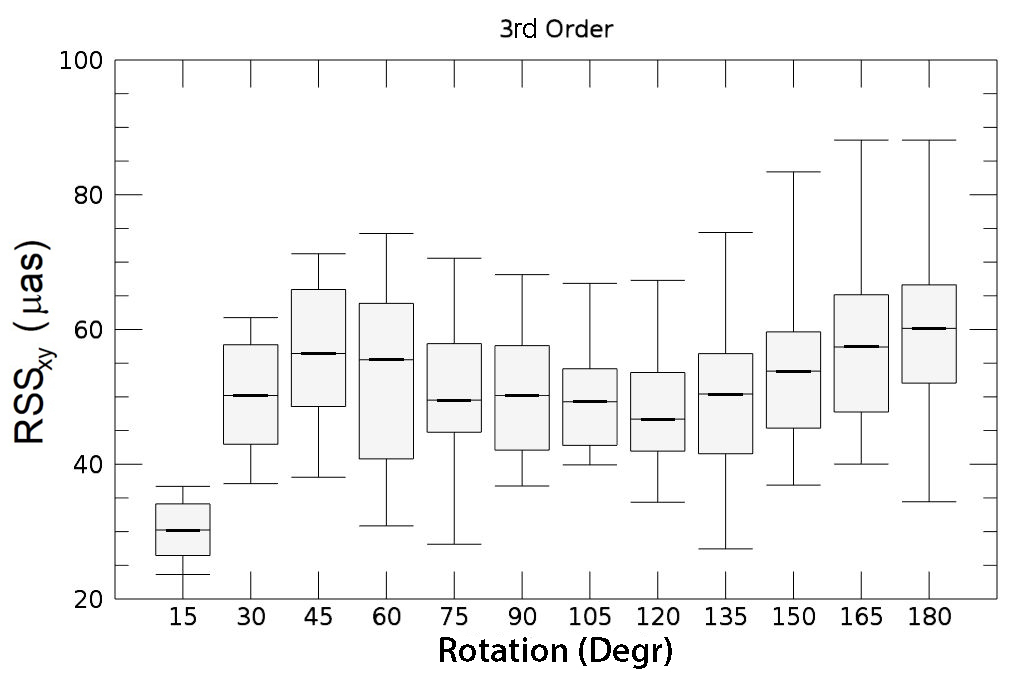}
	\includegraphics[width=0.9\columnwidth]{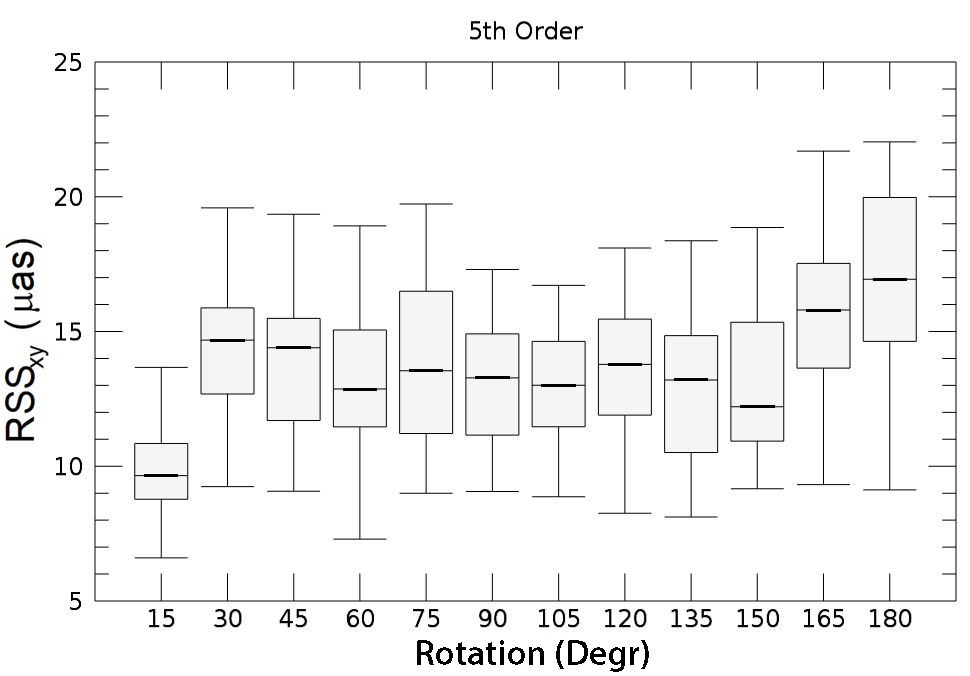}
    \caption{Astrometric precision for a range of field rotations over the scientific FoV as defined by equation~\ref{eq:3}. In the Monte Carlo simulation, optical surface irregularities are coupled with instrument instabilities}. The results after the third and fifth order polynomial correction are shown as representative to achieve the requirement and the goal, respectively. Box plots show minimum and maximum values and quartiles of Monte Carlo trials distribution.
    \label{fig:10}
\end{figure}

When considering multi-epoch observation, optical surface irregularities are coupled with instrument instabilities. The same set of 500 Monte Carlo trials has been perturbed with the expected positioning errors of Table~\ref{tab:1}. A range of field rotation angles has been considered to simulate different epochs. This makes the sources light beams to rotate over the optical surfaces. Given a fixed epoch $E$ and a set of reference sources, the distortion solution has been derived considering the same set of sources at epoch $(E +1)$ where a field rotation has been applied. The analysis results are shown in Figure~\ref{fig:10}. Thirteen epochs are considered, corresponding to 15$^\circ$ of field rotation between two consecutive epochs. In this way we cover the full 180$^\circ$ range, since optical distortions are symmetric respect to the vertical axis. Each point is the residual error as defined in equation~\ref{eq:3}, where the reference star coordinates are those at epoch E. The spread of each point of Figure~\ref{fig:10} is due to the manufacturing tolerances since high order surface irregularities introduce not negligible high order residuals on the polynomial fit. Given the values listed in Table~\ref{tab:2}, about 12\% of the residuals are due to the Zernike coefficients in the range $29\le Z\le45$, 33\% of the residuals are due to $46\le Z\le120$, the rest of the residuals is due to $121\le Z\le230$.
The RMS of the Monte Carlo trials for each angle are represented by the height of the box. In particular, this relates to the amplitude of the Zernike coefficients with higher spatial frequency: greater residuals are due to larger RMS irregularities in the highest Zernike coefficients range.
\\Comparing the upper and lower plots in Figure~\ref{fig:10} we see how at least a fifth order polynomial fit is necessary to achieve or approach the goal of 10 $\mu as$ astrometric precision.\\The worst Monte Carlo trial, in terms of PSF drift within each epoch, is very similar to the result shown in Figure~\ref{fig:9}. Thus, when the centroiding precision is added in quadrature to the fifth-order $RSS_{xy}$, which is about 14 $\mu as$ (median values of Figure~\ref{fig:10}), the instrument is able to achieve the requirement of 50 $\mu as$ of astrometric precision. The goal of 10 $\mu as$ of astrometric precision is achieved for field rotations within 15$^\circ$ in $\sim$50\% of the Monte Carlo trials with the fifth-order $RSS_{xy}$.
\section{Conclusions}
This article describes the foreseen optical distortions of the MAORY module, that we take as example of a MCAO system for an ELT class telescope. Actually, we focus on the astrometric precision achievable by the current optical design described in Section~\ref{sec:2}. The future generation of instruments are designed to be placed on the Nasmyth platform of large telescopes and their optics are not co-rotating with the sky during the tracking of science objects. A possible approach to analyse the impact of optical distortions in terms of PSF drift has been addressed in Section~\ref{sec:4} considering the MCAO correction. Using a pure optical design model, it is possible to simulate the expected behaviour in terms of astrometric performances. Optical simulations combined with a Monte Carlo approach revealed the potential effects of opto-mechanical instabilities as well as the impact of the manufacturing process. This approach can be useful to define calibration procedures which could avoid any additional hardware on the instrument reducing the engineering/calibration time during operations. The MAORY-MICADO system on the ELT has challenging astrometric requirement to be met and shall achieve 50 $\mu as$ of astrometric precision with the goal of 10  $\mu as$ over the MICADO FoV. The tolerances definition of MAORY has been, initially, dictated by degradation of WF quality rather than optical distortion. However, the analysis we present here, confirms the design validity also in terms of astrometric performance. The drift which enlarges the FWHM of the PSF during a single exposure image is within the requirements (4 $mas$) considering the specified tolerances and maximum expected field rotation. Multi-epoch astrometric observations mostly suffer from the manufacturing process which introduces high order optical surface irregularities. Without considering the atmospheric effects, a third order correction is sufficient to remove the degradation due to instrument instabilities. However, it is noticeable that, when coupled with the expected manufacturing, the astrometric precision after a third-order polynomial fit is not sufficient to achieve the requirement of 50 $\mu as$: in this case at least a fourth (fifth) order polynomial fit is necessary. A straightforward observing strategy foresees the mapping of fourth (fifth) order distortions using available stars on each frame: in this way we fulfill the requirements over the MICADO FoV. In our simulations, around 100 point-like sources have been used, which means that the on-sky correction of distortions would be possible using the self-calibration method or using standard astrometric stars in crowded regions where enough Gaia data are available.\\
We described here the opto-mechanical contribution of a more refined astrometric error budget. The full assessment  will take advantage of our analysis and will consider the distortions introduced by atmosphere, telescope and MCAO residuals.
\section*{ACKNOWLEDGMENTS}       
The authors wish to thank all the MICADO consortium for the fruitful on-going collaboration. We are grateful to the anonymous referee for the review work and valuable comments.




\bibliographystyle{mnras}
\bibliography{Astrometria} 


\bsp	
\label{lastpage}
\end{document}